\documentclass{svproc}
\usepackage{graphicx}
\usepackage{longtable}
\usepackage{array} 
\usepackage[numbers,sort&compress]{natbib}

\usepackage{url}

\newcommand{\ding}[1]{%
  \ifnum#1=115%
  \else
    \ding{#1}%
  \fi
}

\begin{document}
\mainmatter              
\title{A bibliographic view on recurrence plots and recurrence quantification analyses}
\titlerunning{Bibliographic view}  
%
\author{Norbert Marwan\inst{1,2,3}}
\authorrunning{Norbert Marwan} 
%
%
\institute{Potsdam Institute for Climate Impact Research (PIK), Member of the Leibniz Association, Germany\\
\email{marwan@pik-potsdam.de}
\and
Institute of Physics and Astronomy, University of Potsdam, Germany
\and
Institute of Geosciences Potsdam, University of Potsdam, Germany
}

\maketitle              

\begin{abstract}
A bibliographic database containing studies on recurrence plots and 
related methods is analyzed from various perspectives. 
This allows a detailed view of the field's development, showcasing the continuous 
growth in the method's popularity, as well as the emergence, decline, and dynamics of 
topical subjects over time.
Furthermore, the analysis unveils the activity and impact of the different groups, 
shedding light on their collaborative efforts and contributions to the field.

\end{abstract}

\section{Introduction}
Quantitative methods to investigate recurring phenomena in theoretical and 
real world applications has become popular in the last decades. One promising
and successful framework is based on the recurrence plot \cite{eckmann87,marwan2007},
a tool to visually represent the temporal recurring patterns of a phase
space trajectory of a dynamical system. It has its roots in nonlinear dynamics
and complex systems \cite{marwan2008epjst}, but extensions and further 
methodological developments were also contributed from other quantitative 
scientific fields. Quantitative extensions include recurrence quantification
analysis \cite{zbilut92,webber94,marwan2002herz}, now a major foundation in this 
analytical framework, as well as
recurrence networks \cite{marwan2009b,donner2010b,zou2019}, an example of 
fruitful combination of different fields
(i.e., recurrence analysis with complex networks).
Further methods and concepts very similar to recurrence plots have been 
developed independently, e.g., recurrence period density entropy \cite{little2007}.

The success of the recurrence plot based framework has been demonstrated by
an impressing collection of exciting applications in different scientific disciplines \cite{marwan2008epjst,marwan2023},
ranging from classical applications in life science, like analysing 
cardiac diseases \cite{wessel2001,iwaniec2018,calderonjuarez2023d} and neurological
disorders \cite{acharya2011a,ngamga2016,billeci2018b}, cognitive sciences
and psychology \cite{riley2003,wallot2013,beimgraben2014}, engineering 
\cite{vlahogianni2008a,mosdorf2012,godavarthi2018,stender2019b,syta2023}, to Earth
sciences \cite{cermak2008b,eroglu2016,oberst2018b,braun2023}.

Since more than 20 years, we maintain a project to get a bibliographic overview
on the methods of recurrence plots, recurrence networks,
recurrence quantification analysis, and related concepts \cite{rpwebsite_bibliography}.
Examining bibliographic data from various perspectives can enhance our 
comprehension of the historical evolution of a scientific field and aid in shaping 
future directions \cite{singh2022bibliographic}. In this study, 
the bibliographic data available at \cite{rpwebsite_bibliography} is analysed, 
to define the different fields where the recurrence plot framework found
applications, to find communities of collaborations and different disciplines,
and to understand the evolution of the different techniques and concepts
which are now a strong integral part of the methodological framework.

\section{Database}\label{sec_database}

\begin{figure}[b!]
\centering
\includegraphics[width=.8\textwidth]{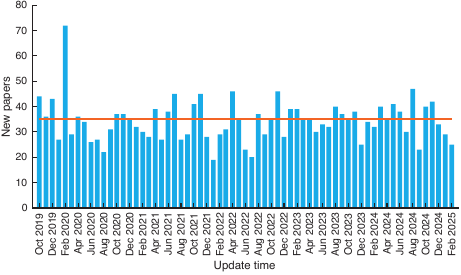}
\caption{Monthly database updates in the last five years. The orange line indicates
the median (34 papers per month).}
\label{fig_databaseUpdates}
\end{figure}

A database of scientific papers using recurrence plots, recurrence quantification
analyses, and recurrence networks is curated by the author since more than 20 years (Tab.~\ref{tab_database}).
Unfortunately, an exact start of the database cannot be reconstructed anymore,
but the Internet Archive reveals one of the first versions 
in 2003 with 125 references \cite{wayback2003}. The database is a 
BibTeX file. It includes articles, conference proceedings, books and book 
contributions, a few PhD thesis, 
as well as some of the available software, in total 4778 entries (by February 2025). 
Among them, the database includes
a few papers which are related to recurrence features in dynamical systems
or alternative approaches looking at recurrences, but not using
the above mentioned methods which have their roots in the work by Eckmann et al. \cite{eckmann87}. 
In the following analysis, these related 
papers and the software entries are neglected, as well as all entries before 1987.
Papers in press are considered as published in 2025, although the final publication
year could still be different.
The resulting bibliographic data set has size $N=4563$.
It is likely that it misses a few publications which are not easy to find.

The database is monthly updated. The update is based on citation alerts
from Clarivate's {\it Web of Science} and Elsevier's {\it Scopus} services and
manual curation. In the last six years, on average 35 new studies join the
database per month (Fig.~\ref{fig_databaseUpdates}).
The data used in this analysis is as of the February 2025 update of the data.

\begin{table}[b!]
\caption{Size and components of the database (excluding papers before 1987 and entries classified as
``related'' and ``software'').}\label{tab_database}
\begin{center}
\begin{tabular}{lr}
\hline
Total number of authors	&9272\\
Total number of papers	&4778\\
Papers analysed			&4563\\
Journal papers (incl.~preprints)	&3656\\
Proceedings				&712\\
Books chapters (ed.)	&174\\
Books					&22\\
Thesis (Bachelor, Master, PhD)	&33\\
Techreports				&4\\
\hline
Number of different journals	&1855\\
Total number of citations&31,118\\
Number of authors' affiliations	&3153\\
Countries of authors' affiliations	&101\\
\hline
\end{tabular}
\end{center}
\end{table}

For the citation analysis, the services of {\it Crossref} and {\it Semanticscholar}
are both utilised to retrieve the cited works within each paper. 
While both services are valuable, they can occasionally introduce slight uncertainties, 
for example, some references may be missing or incorrect in a few papers. 
It has also been observed that following a new run of the {\it Crossref}
retrieval script, certain references were no longer included.
To mitigate these issues, both services are used simultaneously.
However, minor errors may still persist despite these efforts.

Affiliations for all authors of each paper are retrieved from {\it Scopus}. Here, 
not all affiliations could be automatically retrieved and are, thus, missing 
(401 missing from 4563 records, corresponding to 9\% of the papers in the database). Furthermore, 
some minor errors have been found and corrected. For example,
if there were obvious
mistakes or differing names for the same institution, 
these were consolidated (e.g., 
``Rush Medical College'' and ``Rush University Medical Center''
were renamed to ``Rush University'', ``Loyola University Stritch School 
of Medicine'' and ``Loyola University Medical Center'' to
``Loyola University Chicago'', and ``Leibniz-Gemeinschaft'' to
``Potsdam-Institut für Klimafolgenforschung''). 
Nevertheless, despite missing a few affiliations and the possibility of encountering 
some minor glitches in institution names, the retrieved affiliations are deemed 
to be representative of the entire database.

Author names in the database comprise only the surname and the initials of the first name. 
This can pose challenges, as some authors may appear in the database with slightly different 
name variations (e.g., with or without a middle initial), or author names in the database 
may belong to different individuals (particularly with Chinese names, which present a major challenge). 
While some of these issues can be resolved, it may not be possible to address them comprehensively 
across all cases.

The corresponding
scripts for these purposes are available on the mentioned link in
the Data availability section.

\section{Bibliographic analysis}

\subsection{General overview}

\subsubsection{Publications per year.}

\begin{figure}[t!]
\centering
\includegraphics[width=.8\textwidth]{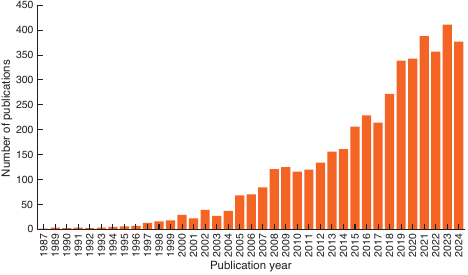}
\caption{Publications per year.}
\label{fig_publicationsPerYear}
\end{figure}

Although very similar concepts like recurrence plots existed before
the seminal paper by Eckmann et al. \cite{eckmann87}, we consider here
this paper as the first one and do not look further back 
(Fig.~\ref{fig_publicationsPerYear}). Including
the current year, we have 38 years of publications (but last year 
not shown). In the first ten years,
only a few papers were published. In 1997, more than ten papers appeared
for the first time in one year. Now, the numbers increased quite quickly,
with reaching the milestone of more than 100 papers per year again almost
ten years later (in 2008), 200 in 2015, and 300 in 2019. After 2019,
the almost exponential grow in publication numbers stoped. Only a slight
increase can be found for the last four years. This is a bit surprising,
as we will see later (Subsect.~\ref{sec_clusters}), a new community has discovered recurrence plots
and is making great use of this method and publishing a lot (hence,
expecting a more rapid increase).

\begin{figure}[t!]
\centering
\includegraphics[width=.8\textwidth]{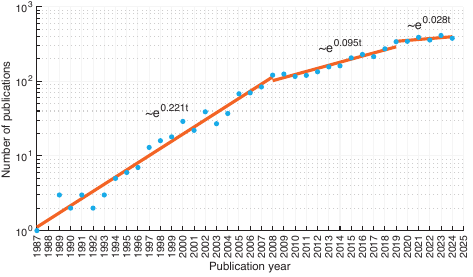}
\caption{Exponential growth of publications over the years (not a cumulative sum). The exponential
growth between 1987 and 2008 was $\sim e^{0.221t}$. Between 2008 and 
2019 and between 2019 and 2024 the growth changed to $\sim e^{0.095t}$ and
$\sim e^{0.028t}$, respectively (orange lines).}
\label{fig_publicationsPerYear2}
\end{figure}

When delving deeper into the details of the exponential growth, 
it becomes evident that the rapid expansion of the field began to slow down 
significantly earlier, around 2008 (Fig.~\ref{fig_publicationsPerYear2}).
The rate of growth was then more than halved. The next change
occurred around 2019, when the growth rate was thirded.
The larger growth rate in the first years (until around 2008/2009) can
be understand by the rapid methodological developments and
introduction of novel concepts, like recurrence quantification analysis 
\cite{webber94,marwan2002herz}, time-dependent recurrence analysis \cite{trulla96},
cross and joint recurrences \cite{zbilut98,romano2004}, interpretation
of line structures and curved patterns \cite{marwan2005,facchini2005},
or recurrence networks
\cite{marwan2009b,donner2010b}. From a methodological point of view, 
the last 15 years have witnessed less fundamentally
new concepts -- although they do exist (e.g., \cite{angus2012,yang2014,corso2018,braun2021,hirata2021a}) -- but rather a proliferation of 
consolidations and robust mathematical underpinnings of the method,
which obviously need more time to be worked out, e.g., \cite{ramdani2016,spitalsky2018,medrano2021,hirata2023a}.

\subsubsection{A growing community.}

\begin{figure}[t!]
\centering
\includegraphics[width=.8\textwidth]{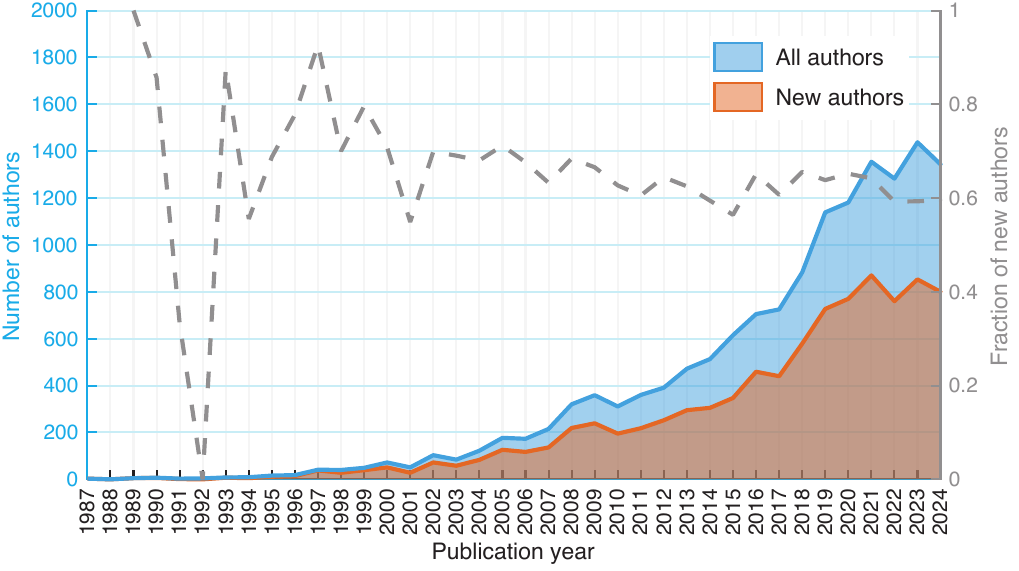}
\caption{Number of authors and new authors per year, as well as the
fraction of new authors.}
\label{fig_numAuthorsPerYear}
\end{figure}

Similar to the number of publications per year, the number of authors per year
and the number of new authors per year are important indicators for
the interest and acceptance of a method. The number of authors and
new authors per year is steadily increasing until 2021 (Fig.~\ref{fig_numAuthorsPerYear}).
In the last three years, these numbers seem to confirm some kind
of saturation within the field, although we still see new scientists
working with recurrence plots and related methods. For most of the time,
the fraction of new authors is mainly in the range between 60 and 70\%, with strong deviations 
in the first few years and a remarkable peak in 1997, when the interest
obviously started to grow rapidly. In the last two years the interest seems
to decrease a bit and the fraction of new authors dropped to 60\%.

\begin{figure}[t!]
\centering
\includegraphics[width=.8\textwidth]{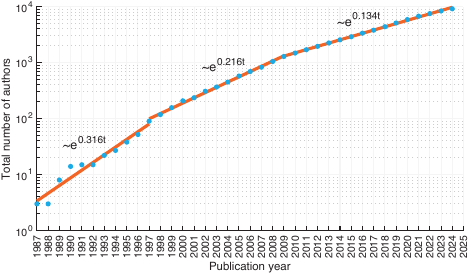}
\caption{Size of the community (total number of authors) on recurrence plot related methods,
counted as the cumulative sum of the new authors. The exponential growth
is changing around 1997 and 2009, with $\sim e^{0.316t}$, $\sim e^{0.216t}$,
and $\sim e^{0.134t}$ (orange lines).}
\label{fig_growingCommunity}
\end{figure}

The cumulative sum of the new authors over the years corresponds to
the size of the scientific community working with recurrence plots and
related methods (Fig.~\ref{fig_growingCommunity}).
In the first six
years, the community is quite small with less than 20 scientists.
After the sixth year, the community grows exponentially, with extending
the size 100 by 1998, and 1000 by 2008 (by February 2025, we have 9272 authors). In general,
this exponential growth could be expected to be similar to the number
of published works. However, we find different change points when comparing
to the number of publications. 
In the first ten years, the community experiences its fastest growth. 
After around 1997, the growth rate declines and after 2009, it further decreases.
This latter
change point corresponds more or less to the change of the exponential
growth in the number of publications and the possible reasons as explained 
above.

\subsubsection{Journals publishing in the field.}

\begin{table}[b!]
\caption{Top journals and selected high prestigious journals publishing in the recurrence plot field,
with total number of publications and fraction (in \%) on all published works.}
\begin{center}
\begin{tabular}{clcc}
\hline
&\textbf{Journal}&\textbf{Publications}	&\textbf{Fraction}\\
\hline
1	&Chaos	&166	&3.79\\
2	&Physical Review E	&82	&1.87\\
3	&Chaos, Solitons \& Fractals	&67	&1.53\\
4	&International Journal of Bifurcation and Chaos	&58	&1.32\\
5	&Physica A	&56	&1.28\\
6	&Physics Letters A	&53	&1.21\\
6	&Scientific Reports	&53	&1.21\\
7	&Lecture Notes in Computer Science	&50	&1.14\\
7	&European Physical Journal -- Special Topics	&50	&1.14\\
8	&Entropy	&48	&1.10\\
9	&Nonlinear Dynamics	&43	&0.98\\
9	&PLoS ONE	&43	&0.98\\
9	&Sensors	&43	&0.98\\
10	&Biomedical Signal Processing and Control	&35	&0.80\\
11	&Frontiers in Psychology	&34	&0.78\\
12	&IEEE Access	&29	&0.66\\
13	&IEEE Sensors Journal	&26	&0.59\\
14	&Europhysics Letters	&25	&0.57\\
	&\vdots \\
28	&Proceedings of the National Academy of Sciences		&7	&0.16\\
28	&Physical Review Letters	&7	&0.16\\
29	&Heliyon	&6	&0.14\\
29	&Nature Communications	&6	&0.14\\
31	&iScience	&4	&0.09\\
32	&Science Advances	&3	&0.07\\
34	&Nature		&1	&0.02\\
34	&Science	&1	&0.02\\
\hline
\end{tabular}
\end{center}
\label{tab_topjournals}
\end{table}

The top journals which publish studies using recurrence plot based methods
are mainly from the fields of physics, in particular on 
nonlinear and complex systems science,
such as Chaos, Physical Review E, Physica A, Chaos, Solitons \& Fractals,
and International Journal of Bifurcation and Chaos
(Tab.~\ref{tab_topjournals}). Among the top journals
are also some from more specific scientific disciplines such as
biomedical and psychological sciences, and, a bit surprising, from a more
engineering related field on applications of sensors: 
Biomedical Signal Processing and Control,
Frontiers in Psychology, (MDPI) Sensors, and IEEE Sensors.
PLoS ONE, Scientific Reports, and IEEE Access represent the only
general journals in this top list, although
others high prestigious journals have published studies using recurrence
plot based methods, like
Proceedings of the National Academy of Sciences (7), Physical Review Letters (7),
Heliyon (6), and Nature Communications (6), or even 
Nature and Science (1).

\begin{figure}[t!]
\centering
\includegraphics[width=.95\textwidth]{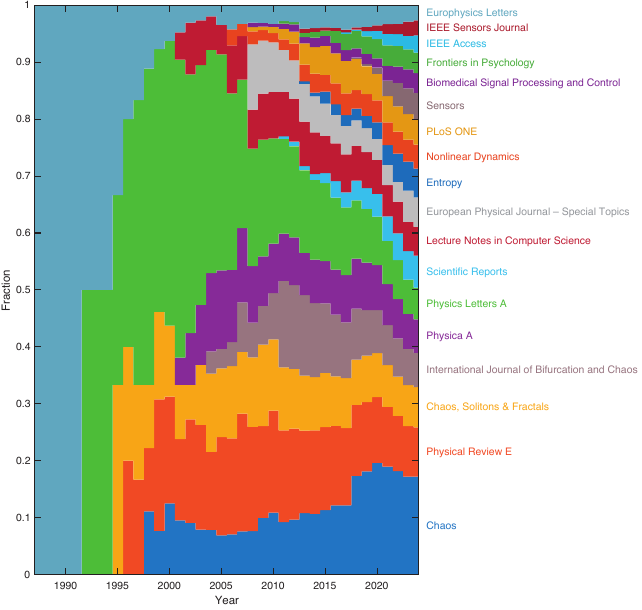}
\caption{Evolution of the distribution of the top journals from 1987 to 2024
where the assorted colors are stacked upon one another with no hidden data and 
each year sums to 100\%.}
\label{fig_journalsPerYear}
\end{figure}

The seminal paper on recurrence plots by Eckmann et al. was published in
Europhysics Letters (EPL) \cite{eckmann87}. However, this journal 
has not further received much attention from other authors publishing in this
field (Fig.~\ref{fig_journalsPerYear}). Its popularity dropped down to position
14 in the year 2025 (Tab.~\ref{tab_topjournals}). For more than a decade,
Physics Letters~A was the main journal for publishing recurrence plot
papers. Between 2005 and 2015, Physical Review E and
International Journal of Bifurcation and Chaos have published
a lot. Since 2005 (first paper in 1998), Chaos is steadily increasing in the interest of authors.
Since 2008 it is the journal which is publishing most of the works and after 2017
it received a jump in attracting papers on recurrence methods.

Other important journals in this field are, as mentioned before, Physica A,
which has seen first recurrence plot papers in 2001, and
Chaos, Solitons \& Fractals, since 1995.
Lecture Notes in Computer Science and European Physical Journal -- Special Topics
are often used for conference proceedings. Newer journals are PloS ONE,
Scientific Reports, and Entropy. Since 2018, the number of papers in Sensors is rapidly 
growing, mainly with papers on recurrence analysis in combination with machine learning.
The growth of journals like Frontiers in Psychology and Biomedical Signal Processing and Control
in the field is an indication for a strong community in these disciplines
which uses recurrence methods and that these methods have become established methods
there.

\subsection{Subject analysis}\label{sec_clusters}

\begin{figure}[b!]
\centering
\includegraphics[width=.8\textwidth]{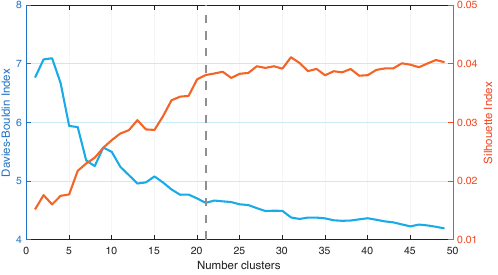}
\caption{Davies-Bouldin and silhouette score for increasing number of clusters
during GMM clustering of the papers. The dashed line indicates the local minimum
in the Davies-Bouldin index and the selected cluster number of 21.}
\label{fig_DBindex}
\end{figure}

In order to get an overview on the different scientific disciplines,
recurrence plot based methods are used in, subjects for
each paper in the database should be assigned. Using a service like {\it Crossref} has
not been useful for this purpose. The resulting subjects are too much overlapping
and the resulting clusters of papers are not informative enough.
An alternative would be a clustering analysis of the titles, abstracts,
and keywords (when available). Here, a GMM (Gaussian mixture model clustering) clustering is applied.
The words from the titles, abstracts, and keywords are merged for each paper 
and automatically and manually transformed into their base or dictionary form, reducing each word to 
its root form to facilitate analysis and comparison (lemmatization).

To find the optimal number of clusters, the criteria are that
the clusters should not overlap too much (which
is quite difficult when using titles and abstracts) but reflect
meaningful subject-based clustering. The Davies-Bouldin index \cite{davies1979}
and silhouette score \cite{rousseeuw1987} are used to 
numerically support the choice of an optimal cluster number. 
The Davies-Bouldin index measures the similarity between the clusters
and the silhouette score evaluates how well-separated the clusters are.
The Davies-Bouldin index indicates a better clustering
with lower scores, whereas the silhouette score should have larger scores. 
Increasing the number
from 1 to 50 clusters reveals a rapid decrease of the Davies-Bouldin index
until a cluster number of 13 (Fig.~\ref{fig_DBindex}). 
After this cluster number, the index decreases slower, with a small local
minimum at 22.
The silhouette score shows an almost linear increase up to 20, 
after which it levels off into a plateau.
A cluster number of 21 seems quite appropriate and is a trade-off to have 
enough subject-specific clusters and only a low number of
redundant clusters.

This analysis groups the papers into different subjects (Tab.~\ref{tab_clusters}). The result
is not perfect. Some clusters contain more than one subject (e.g., cluster 6 contains some 
more general time series analysis studies, but also papers on vibration monitoring), 
or even some subjects are missing or hidden
within the dominating topic of a cluster. 
However, for a larger number of
clusters, many redundant clusters appear, i.e., with very similar
topics. Similar topics are already visible for the choice of 21 clusters (for larger cluster numbers
the number of redundant cluster increases significantly), such as
clusters 1, 3, 5, and 6; 3, 6, and 14; 3 and 17; 8 and 20; or 12 and 16. 

\setlength{\tabcolsep}{7pt} 

{\small
\begin{longtable}{c>{\raggedright\arraybackslash}p{3cm}c>{\raggedright\arraybackslash}p{6.cm}}

\caption{{\normalsize Clusters based on the titles, abstracts, and keywords of the paper
in the database. Based on the 15 top features (words), subjects have been assigned
to the clusters.}}\label{tab_clusters}\\
\textbf{Cluster} & \textbf{Subject} & \textbf{Papers} & \textbf{Top words} \\
\hline
\endfirsthead

\textbf{Cluster} & \textbf{Subject} & \textbf{Papers} & \textbf{Top words} \\
\hline
\endhead

\hline
1    &Nonlinear Time Series Analysis   &586    &timeseries, data, recurrenceplot, dynamic, recurrence, nonlinear, dynamical, application, time, complex, recurrencequantification, model, information, structure, measure\\
2    &Social Interaction Dynamics    &364    &coordination, interaction, movement, task, social, eye, participant, interpersonal, team, synchrony, human, behavior, dynamic, recurrencequantification, cross\\
3    &Signal Detection in Monitoring Tasks &360    &signal, detection, vibration, recurrenceplot, monitoring, damage, recurrencequantification, feature, frequency, recurrence, technique, time, power, condition, algorithm\\
4    &Recurrence Plots as Features for Machine Learning    &354    &image, neuralnetwork, classification, convolutional, cnn, deeplearning, feature, accuracy, model, data, recurrenceplot, recognition, timeseries, deep, signal\\
5    &Nonlinear Dynamical Systems   &340    &chaos, dynamic, exponent, lyapunov, attractor, nonlinear, dynamical, bifurcation, periodic, timeseries, recurrenceplot, model, motion, transition, dimension\\
6	 &Nonlinear Feature Extraction	&276	&recurrence, corrosion, recurrenceplot, noise, process, recurrencequantification, parameter, entropy, determinism, value, dynamic, line, surface, rate, current\\
7    &Biomechanics, Motor control, and Neuromuscular Physiology   &247    &muscle, postural, control, emg, surface, group, subject, recurrencequantification, motor, healthy, determinism, balance, movement, signal, nonlinear\\
8    &Cardiovascular Physiology   &240    &heart, rate, variability, hrv, cardiac, autonomic, nonlinear, patient, ecg, cardiovascular, interval, signal, time, recurrenceplot, subject\\
9    &Brain Activity and Cognitive Science   &215    &brain, eeg, activity, functional, state, signal, cognitive, subject, recurrence, patient, dynamic, nonlinear, measure, potential, network\\
10   &Epileptic Seizure Detection   &208    &eeg, feature, classification, signal, seizure, epileptic, classifier, accuracy, svm, detection, recurrencequantification, nonlinear, disorder, extracted, patient\\
11	 &Studies in Climate, Earth science, and Astrophysics	&197		&climate, solar, temperature, record, cycle, change, variation, period, year, long, event, dynamic, oscillation, time, activity\\
12   &Fluid Flow Characteristics in Pipelines   &166    &flow, gas, bed, phase, water, pressure, velocity, pattern, fluctuation, oil, recurrenceplot, recurrence, signal, characteristic, dynamic\\
13   &Recurrence Networks	&151	&network, recurrencenetwork, complex, timeseries, dynamical, property, dynamic, measure, graph, phasespace, recurrence, nonlinear, chaos, data, world\\
14	 &Classifying Dynamical System Types		&150		&deterministic, stochastic, chaos, dynamic, timeseries, nonlinear, data, process, recurrenceplot, model, noise, recurrencequantification, signal, component, dynamical\\
15	 &Synchronisation and Transition Analysis	&145	 &synchronization, phase, coupling, coupled, oscillator, network, recurrence, chaos, cross, transition, time, state, probability, nonlinear, different\\
16    &Combustion Processes \& Thermoacoustic Instability   &132    &combustion, instability, thermoacoustic, flame, oscillation, pressure, acoustic, dynamic, transition, flow, cycle, amplitude, ratio, fluctuation, rate\\
17    &Fault Diagnosis in Mechanical Systems   &125    &fault, diagnosis, bearing, vibration, signal, recurrenceplot, feature, neuralnetwork, image, condition, model, accuracy, convolutional, detection, data\\
18    &Financial Market Analysis   &108    &market, price, financial, timeseries, index, chaos, recurrenceplot, dynamic, data, energy, recurrencequantification, time, behavior, nonlinear, correlation\\
19    &Protein Structure \& Folding Analysis   &80    &protein, sequence, structure, structural, recurrencequantification, prediction, class, feature, property, region, interaction, recurrence, information, set, function\\
20    &Atrial Fibrillation Prediction   &72    &atrial, af, fibrillation, ecg, patient, recording, pattern, signal, recurrence, length, feature, recurrencequantification, cycle, clinical, cardiac\\
21	  &Monitoring and Prediction in (Road and IT) Traffic Networks	&47	&traffic, network, flow, anomaly, detection, statistical, data, short, term, prediction, detect, dynamic, recurrence, recurrencequantification, recurrenceplot\\
\hline
\end{longtable}
}

The largest cluster (cluster 1) is on the subject of nonlinear time series
with 586 papers, including the seminal paper by Eckmann et al.~\cite{eckmann87}.
The second largest cluster (cluster 2) is on dynamics of social interactions. 
Several of the larger clusters (7 to 10) are related to life sciences 
(psychology, cognitive sciences, cardiovascular and brain physiology).
This shows that there is solid community
in these fields working with recurrence plot based methods.
Similarly, engineering-related topics are spread over several clusters (3, 6, 12, 16, 17, 21).
We further find clusters on financial markets and protein folding (clusters 18 and 119).
Not to forget the rather new but rapidly growing field of machine learning-based classification approaches, which utilize 
recurrence properties as features in learning-based classification methods (cluster 4).

\begin{figure}[h!]
\centering
\includegraphics[width=.95\textwidth]{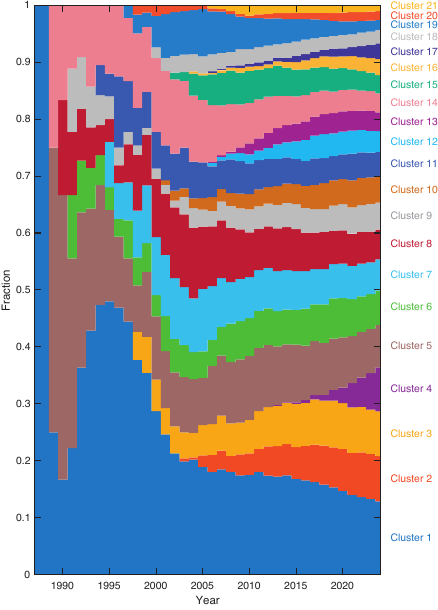}
\caption{Subjects of papers evolving over the years (starting point
of time interval is always 1987). The assorted colors are stacked upon 
one another with no hidden data and each year sums to 100\%.}
\label{fig_clustersPerYear}
\end{figure}

When applying the same clustering schema to a subset of the papers restricted by time
intervals (from 1987 to a specific year), it is possible to study how selected subjects
have evolved over time, when specific achievements were reached, or when hot topics 
emerged (Fig.~\ref{fig_clustersPerYear}).

Since the method is nonlinear time series analysis and has its roots in the fields chaos
and dynamical systems, it comes as no surprise that clusters 1 and 5 are among 
the initial ones to emerge.
Studies on cardiovascular and brain dynamics (clusters 7 and 8) were the first 
applications of this method; corresponding clusters start in 1990 and 1991. Next applications
were found in neuromuscular physiology (1995) and protein structure analysis (1997).
Interestingly, the latter subject peaked between 2000 and 2009, but gradually declining thereafter and 
showing a relative decrease in its community size within this field (in total numbers it is still increasing).
Additional subjects can be identified along with their inception dates, e.g., 
recurrence network analysis (cluster 13), starting with 2006 and abruptly increasing in 2009, 
combustion analysis (cluster 16), strongly increasing around 2015,
and early works on solar physics (cluster 11), starting 1994.
Notably, cluster 4, focusing on feature-based classification and machine learning, 
has experienced rapid growth since 2014, signifying the onset of the machine learning hype after this year.
Applications in engineering (clusters 12, 16, 17, 21) have started and expanded after the mid-2000ies.

\subsection{Citation analysis}

Paper citations serve as a measure of scholarly influence \cite{portenoy2017}.
For such analyses, the references of all papers in the database are collected
from {\it Crossref} and {\it Semanticscholar} (see Sec.~\ref{sec_database})
and used to find the citations of papers. This approach ensures
that we only consider in-community citations and the influence of papers
within the recurrence plot/ recurrence quantification field. 
In general, a paper may have received many more additional citations beyond those analysed here.
Moreover, due to the found errors and inconsistencies in the 
{\it Crossref} and {\it Semanticscholar} database, the number might not be
100\% accurate.
In total, the papers in the database have cited each other 31,118 times.

\begin{figure}[b!]
\centering
\includegraphics[width=.8\textwidth]{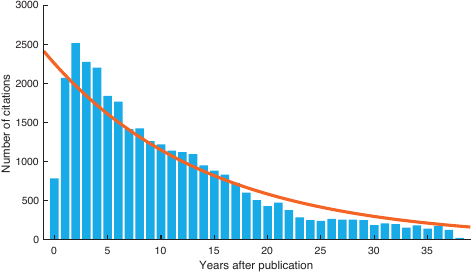}
\caption{Time between publication of a paper and its citations. The exponential decay
is fitted as $e^{-0.07t}$. Only citations within the RP/RQA community are considered.}
\label{fig_citationTimes}
\end{figure}

The citation time of a paper, defined as the duration between its publication 
and when it garners citations, can serve as a metric for assessing the 
topicality of a subject. Papers that are frequently cited shortly after 
publication, typically within the same year or shortly thereafter, 
indicate a `hot topic' or a high level of relevance and significance \cite{yan2010a}.
In the recurrence plot database, papers were mostly cited after two to
four years (Fig.~\ref{fig_citationTimes}). After this time, the number
of citations decreases the older the paper is. This decrease follows
approximately an exponential decay, $e^{-0.07t}$. Only the most relevant papers
are further cited.

The papers which were most frequently cited within the same or the next year
are the review papers \cite{marwan2007,zou2019,donner2010b} 
and the edited book \cite{webber2015} (Tab.~\ref{tab_citationrank}). 
These papers are also the ones which are cited for longer times.
Among these most frequently cited papers within in the first year are 
further the topical papers on recurrence networks \cite{donner2010b,donner2010a,donner2011,donges2012}
and on combining machine learning with recurrence approaches \cite{mathunjwa2021,zhang2022}.
These examples express well the very topical nature of these subjects.

\begin{figure}[t!]
\centering
\includegraphics[width=.8\textwidth]{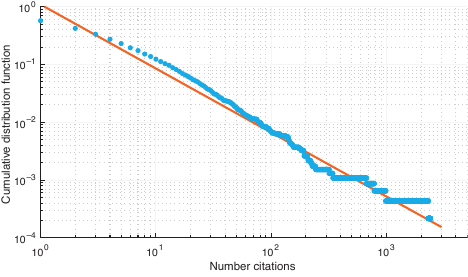}
\caption{Cumulative distribution function of citation numbers (papers
not cited are not visible in this plot, because of the log-log
axes). The scaling exponent of the fit is $\sim N^{-1.1}$ (with $N$ the number of citations).
Only citations within the RP/RQA community are considered.}
\label{fig_citationNumbers}
\end{figure}

Of the 4563 papers, 43\% ($n=1971$) were not cited at all
and about 57\% were cited at least once (Fig.~\ref{fig_citationNumbers}). 
12\% of the papers were cited at least 10 times, 6\% were cited 20 times
and 0.7\% were cited 100 times. The top ten papers in this ranking
have more than 200 citations. Among them, 
the two seminal papers \cite{eckmann87,marwan2007}
received about 2200 citations each (Tab.~\ref{tab_citationrank}).

\begin{table}[t!]
\caption{Papers cited more than 200 times ($C_\mathrm{all}$) and 
more than 9 times during the first two years after publication
($C_2$), considering only citations by the other papers in the
recurrence plot database (ranked regarding the overall number of
citations). Only citations within the RP/RQA community are considered.}
\begin{center}
\begin{tabular}{rlrrc}
&\textbf{Paper ID} 	&\textbf{C$_\mathrm{all}$}	&\textbf{C$_2$}	&\textbf{Reference}\\
\hline
1	 &marwan2007	&2404	&68	&\cite{marwan2007}\\
2	 &eckmann87	&2297	&0	&\cite{eckmann87}\\
3	 &webber94	&985	&1	&\cite{webber94}\\
4	 &zbilut92	&782	&1	&\cite{zbilut92}\\
5	 &marwan2002herz	&669	&7	&\cite{marwan2002herz}\\
6	 &marwan2002pla	&342	&6	&\cite{marwan2002pla}\\
7	 &trulla96	&318	&0	&\cite{trulla96}\\
8	 &marwan2009b	&242	&7	&\cite{marwan2009b}\\
9	 &zbilut98	&226	&0	&\cite{zbilut98}\\
10	 &marwan2011	&220	&8	&\cite{marwan2011}\\
11	 &marwan2008epjst	&208	&3	&\cite{marwan2008epjst}\\
12	 &donner2010b	&206	&15	&\cite{donner2010b}\\
20	 &donner2011	&144	&13	&\cite{donner2011}\\
21	 &webber2015	&143	&12	&\cite{webber2015}\\
29	 &zou2019	&100	&29	&\cite{zou2019}\\
48	 &donges2012	&59	&10	&\cite{donges2012}\\
52	 &donner2010a	&54	&13	&\cite{donner2010a}\\
74	 &mathunjwa2021	&30	&11	&\cite{mathunjwa2021}\\
87	 &marwan2023	&17	&15	&\cite{marwan2023}\\
87	 &zhang2022	&17	&10	&\cite{zhang2022}\\
\hline
\end{tabular}
\end{center}
\label{tab_citationrank}
\end{table}

\begin{figure}[t!]
\centering
\includegraphics[width=.8\textwidth]{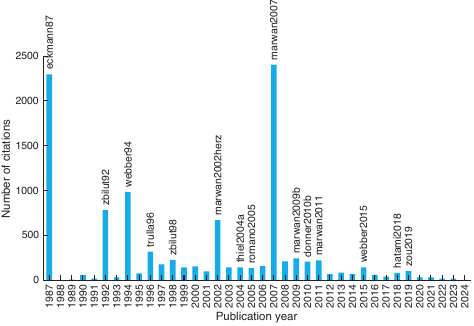}
\caption{Maximum number of citations reached by a paper of a selected publication year,
indicating milestone works. Only citations within the RP/RQA community are considered.}
\label{fig_citationJumps}
\end{figure}

For each year, we can identify the paper with the highest number of citations. 
Considering these data over time reveals peaks that highlight milestone
works in the field (Fig.~\ref{fig_citationJumps}). Given that more 
recent papers tend to have fewer citations, we can identify the following
significant milestones. 

The first paper on recurrence plots by Eckmann et al. in 1987~\cite{eckmann87}
is, of course, one with the highest peaks. Following that, the introduction 
of the recurrence quantification idea in 1992 by Zbilut and Webber, 
along with the addition of additional measures in 1994 
\cite{zbilut92,webber94}, represent the next significant peaks. The approach
of windowed recurrence quantification analysis by Trulla et al.~in 1996
\cite{trulla96} is also discernible. A small peak in 1998 marks
the introduction of the cross-recurrence idea \cite{zbilut98}.
The next notable peak appears in 2002 with the
introduction of the recurrence quantification measures based on vertical lines 
\cite{marwan2002herz}. In 2004 and 2005, small peaks emerge, associated with
studies on estimating dynamical invariants from recurrence plots \cite{thiel2004a}
and detecting phase synchronisation with an innovative bivariate extension of
recurrence plots \cite{romano2005}. The highest peak occurs in 2007
when a review paper on the method was published \cite{marwan2007}.
Subsequent peaks in 2009 and 2010 correspond to the
introduction of the recurrence network approach \cite{marwan2009b,donner2010b}.
The peak in 2011 is related to a critical paper highlighting potential pitfalls
associated with the method \cite{marwan2011}. Peaks in 2015 and 2019 are linked
to an edited book on recurrence quantification analysis \cite{webber2015}
and a review paper on network-based time series analysis \cite{zou2019},
respectively. The small peak at 2018 indicates one of the first papers
using RP-based time series imaging for machine learning \cite{hatami2018}.

\subsection{Impact of Institutions}\label{sec_impactInst}

The activity and impact of institutions is represented by the number of publications
and their respective citations. To avoid multiple counting within one paper,
an institution is counted only once if co-authors from a paper 
are from the same institution.
In total, the {\it Scopus} search for the affiliations of all
authors in the database found 3153 different institutions. Unfortunately, the search was not
successful for every paper, thus, the statistics is missing some counts (see Sec.~\ref{sec_database}).
We should also remember that authors
can change their affiliations due to moving to another institute, therefore,
lowering or biasing the presented numbers for specific institutions.
Some authors can also have more than one affiliation, thus contributing
to more than one institution and further biasing the statistics (as, in
particular, happened for the top ranked institutions as we will see below).

\begin{table}[t!]
\caption{Number of publications $P$, number of total
citations $C$, and average citations per paper $A$ of the top ten 
institutions (regarding the number of publications) and 
(additionally) the institutions with more
than 40 publications or more than 500 average citations per paper.
Only citations within the RP/RQA community are considered (the total number
of citations is much higher).}
\begin{center}
\begin{tabular}{rlrrr}
\textbf{} 	&\textbf{Institute} 	&\textbf{P}	&\textbf{C}	&\textbf{A}\\
\hline
1	&Potsdam Institut fur Klimafolgenforschung	&191	&3703	&19.39\\
2	&Humboldt-Universität zu Berlin	&129	&3062	&23.74\\
3	&Universität Potsdam	&109	&6554	&60.13\\
4	&Politechnika Lubelska	&85	&464	&5.46\\
5	&Tianjin University	&82	&397	&4.84\\
6	&Indian Institute of Technology Madras	&72	&523	&7.26\\
7	&University of Aberdeen	&68	&1166	&17.15\\
8	&University of Cincinnati	&60	&691	&11.52\\
9	&Istituto Superiore Di Sanita	&56	&950	&16.96\\
10	&Rush University	&55	&2629	&47.80\\
11	&Sapienza Università di Roma	&51	&693	&13.59\\
12	&Universidade Federal do Parana	&49	&292	&5.96\\
13	&Loyola University Chicago	&47	&3452	&73.45\\
27	&Université de Genève	&3	&2257	&752.33\\
28	&Institut des Hautes Études Scientifiques	&2	&2272	&1136.00\\
\hline
\end{tabular}
\end{center}
\label{tab_institutepapers}
\end{table}

The most active institutions based on number of publications
(considered as the top 10) over the entire
time are mainly related to seven groups: 
the group at the
(1) Potsdam Institute for Climate Impact Research (PIK), Humboldt University Berlin, University of Potsdam,
and University of Aberdeen
(these four institutions are associated with the group of authors around Kurths and Marwan;
where Kurths was previously affiliated with the University of Potsdam and is currently affiliated 
with the Potsdam Institute for Climate Impact Research, Humboldt University Berlin, and the University of Aberdeen), 
(2) the Polytechnical University Lublin (group around Litak),
(3) the Tianjin University (group led by Gao and Jin),
(4) the Indian Institute of Technology Madras (group led by Sujith),
(5) the University of Cincinnati (group led by Kiefer and Riley),
(6) the Istituto Superiore Di Sanita and Sapienza University in Rome (group around Giuliani),
(7) and Rush University and Loyola University Chicago (group around Zbilut and Webber, 
Loyola University Chicago is mentioned here although it is not in the top ten)
(Tab.~\ref{tab_institutepapers}, column P).
The three institutions with high numbers of publications -- 
the Polytechnical University Lublin,
the Tianjin University, 
and the Indian Institute of Technology Madras --
are noteworthy because they are primarily associated with studies in the field of engineering science.

The citation analysis on the institutional level reveals 
the importance and impact of the institutions in this field
(Tab.~\ref{tab_institutepapers}, columns C and A). 
Considering the average number of citations per paper, the 
Institut des Hautes Études Scientifiques and
University of Geneve are the top institutions with more than 750 
(and even 1000) citations per paper. These are the
host institutions of the authors of the first paper on recurrence plots \cite{eckmann87}.

The next top institutions have more than 40 average citations per paper (but less than 100):
Loyola University Chicago, Rush University, and University of Potsdam, with
introduction of and major contributions on the recurrence quantification analysis
\cite{zbilut92,webber94,marwan2002herz},
as well as the review paper on recurrence plots \cite{marwan2007}. Average per paper citations
of more than 10 are still quite remarkable and draw 
the Humboldt University Berlin, 
the Potsdam Institute for Climate Impact Research (PIK),
the University of Aberdeen, 
Istituto Superiore Di Sanita in Rome,
Sapienza University in Rome,
and the University of Cincinnati to the most important institutions in the field.

\subsection{Authors}

\begin{figure}[b!]
\centering
\includegraphics[width=.8\textwidth]{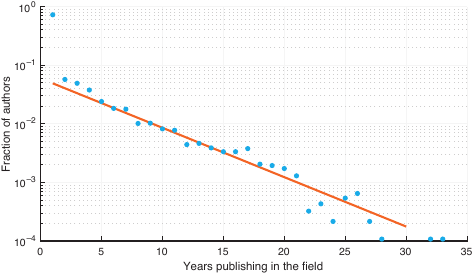}
\caption{Time in years authors are publishing in the field. The exponential fit is 
$e^{-0.19t}$.}
\label{fig_authorsTime}
\end{figure}

The number of author's publications and the duration authors 
remain active in publishing within the field serve 
as indicators of their experience and expertise. The majority of authors
in the database has published only once (69\%) 
and only in one year (73\%; see Fig.~\ref{fig_authorsTime}),
29\% of them have published as first author. 
The distribution of the duration authors are publishing follows
an exponential decay $\sim e^{-0.19t}$.
Less than 0.2\% of authors have published at least 20 years (Tab.~\ref{tab_authorstime}).
Webber and Kurths have been active in the field for over 30 years, 
making them the longest-standing contributors. They are closely followed by Kaplan 
and Giuliani in terms of their duration of activity.

\begin{table}[t!]
\caption{Authors publishing at least for 20 years on recurrence plot based
methods. The duration of publications is based on the final publication date, 
not the submission date, and may include posthumous publications.}
\begin{center}
\begin{tabular}{r>{\raggedright\arraybackslash}p{8cm}c}
\textbf{} 	&\textbf{Author} 	&\textbf{Duration}\\
\hline
1	&Webber, Jr., C. L.	&33\\
2	&Kurths, J.	&32\\
3	&Kaplan, D.	&28\\
4	&Giuliani, A.; Balasubramaniam, R.	&27\\
5	&Schulz, S.; Orsucci, F.; Marwan, N.; Kankaanp\"a\"a, M.; Dong, F.; Airaksinen, O.	&26\\
6	&Zimatore, G.; Schmidt, J.; Riley, M. A.; Hu, Z.; Banerjee, S.	&25\\
7	&Santarcangelo, E. L.; Faure, P.	&24\\
8	&Zbilut, J. P.; Trulla, L. L.; Strozzi, F.; Newell, K. M.	&23\\
9	&Trauth, M. H.; Pereda, E.; Li, X.	&22\\
10	&Wang, X.; Meng, Y.; Matcharashvili, T.; Liu, Y.; Liu, X.; Liu, C.; Litak, G.; Li, J.; Hatzopoulos, S.; Chen, H.; Chelidze, T.; Acharya, U. R.	&21\\
11	&Zou, Y.; Zhang, L.; Yao, X.; Wang, W.; Patel, M.; Lu, S.; Lu, H.; Liu, G.; Li, Y.; Li, M.; Keogh, E.; Huang, Y.; Faust, O.; Czarnigowski, J.; Chen, W.; Cerutti, S.	&20\\
\hline
\end{tabular}
\end{center}
\label{tab_authorstime}
\end{table}

\subsection{Collaborations}

Finally, we consider the collaborations of the scientists in the fields of
recurrence plots and recurrence quantification. 

\subsubsection{Country-level collaborations.}

\begin{figure}[b!]
\centering
\includegraphics[width=1\textwidth]{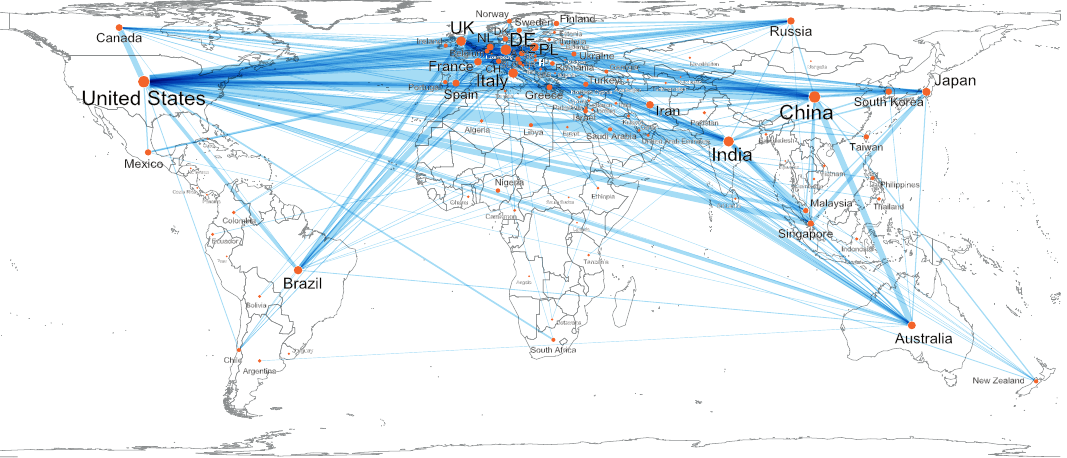}
\caption{Country-level collaboration network of co-authors. Link width indicate number
of joint papers, node size correspond to number of publications assigned to
the country (the larger the more publications).}
\label{fig_affiliationsCountry_network}
\end{figure}

A general view can 
be achieved by looking at the countries of the authors (based on the
affiliations, retrieved by {\it Scopus}). 
We find in total 101 countries with
authors publishing on recurrence plots, where an author can have multiple
affiliations in different countries, and therefore, could be counted several times.
Among these countries are the US, China, Germany,
India, UK, Italy, and Poland the most active countries (with more than 200 publications each).
The collaboration network of the co-authors at the country level 
reveals strong collaborations between UK and Germany, US and Italy,
Germany and US, China and US, UK and US, India and US, and UK and China, to mention
the top collaborations based on the volume of joint papers produced
(Fig.~\ref{fig_affiliationsCountry_network}). We also observe widespread 
usage of the method across all continents, with Africa showing relatively 
fewer contributions and collaborations compared to other regions.

\subsubsection{Co-author network.}

\begin{figure}[b!]
\centering
\includegraphics[width=.8\textwidth]{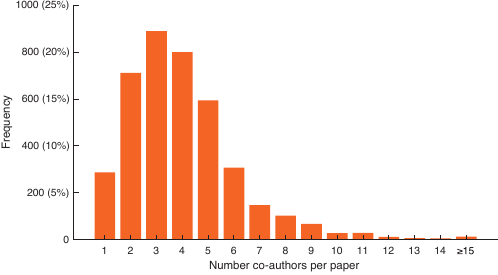}
\caption{Frequency distribution of the number of co-authors, with median $=4$, and 
1$^\mathrm{st}$ and 3$^\mathrm{rd}$ quartile as $Q_1 = 3$ and $Q_3 = 5$ co-authors.}
\label{fig_authorsPerPaper}
\end{figure}

A more detailed analysis can be performed on the co-author level. In average, 
a paper is written by four authors (median). The number of co-authors ranges
from one to a maximum of 40 authors \cite{asghari2004} (Fig.~\ref{fig_authorsPerPaper}).

\begin{figure}[htbp]
\centering
\includegraphics[width=\textwidth]{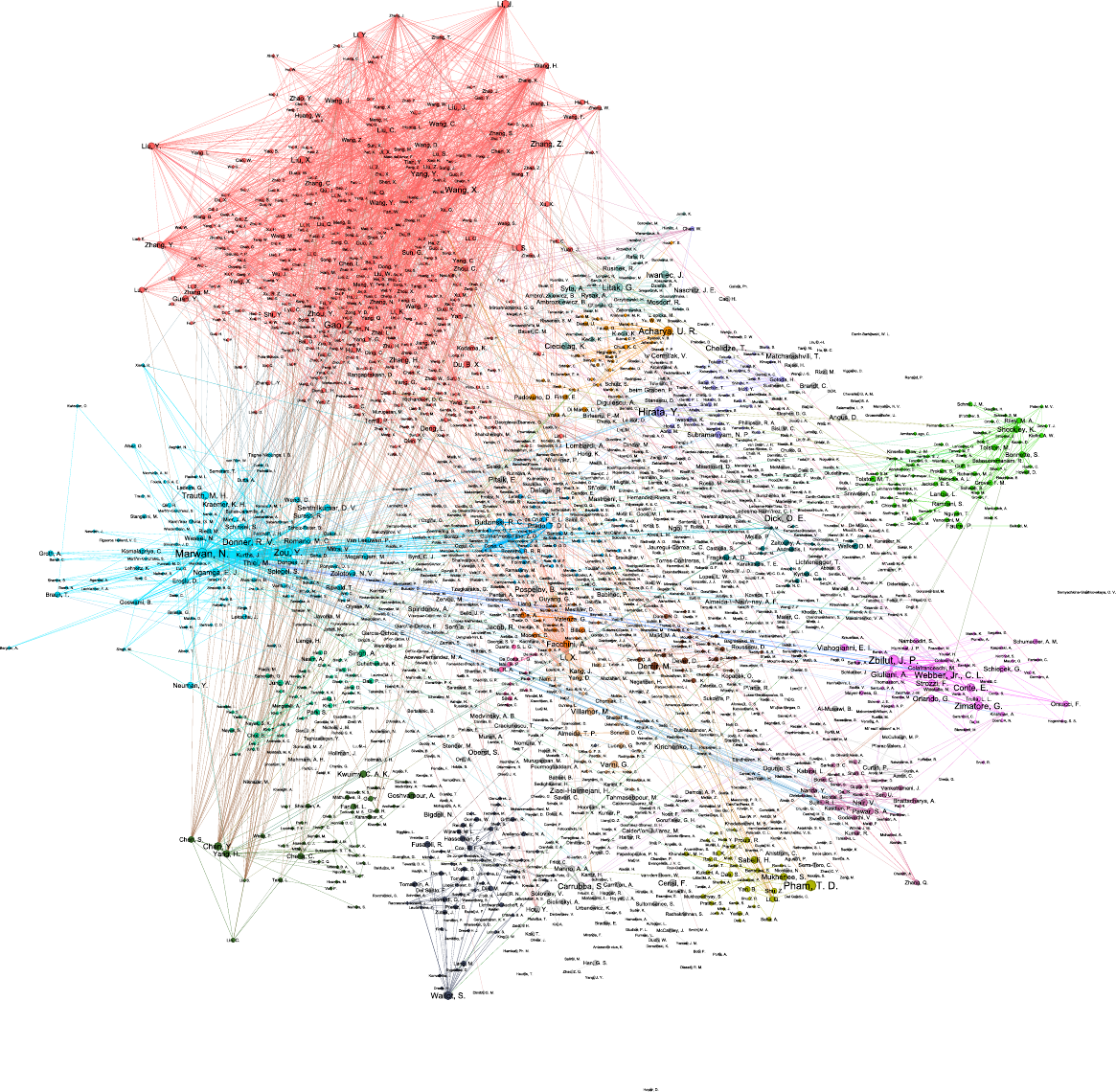}
\caption{Co-author network, filtered to display only first-authors who have engaged in at least 
one collaboration (therefore, some communities appear to be splitted).
Node size corresponds to the number of first-author papers, colour indicates different communities
(with only the largest communities with at least 1\% of authors are being coloured).}
\label{fig_coauthor_network}
\end{figure}

By constructing a co-author network, communities can be easily identified (Fig.~\ref{fig_coauthor_network}).
In general, the communities are formed by groups at universities
(e.g., supervisor and students, project collaborators), and by specific disciplines (e.g., complex systems,
cognitive science, life sciences, engineering). 
After applying a community detection algorithm \cite{blondel2008}, the largest communities (Tab.~\ref{tab_community})
are the ones led by 
Wang, X.; Wang, J.; and Wang, X. (community 574);
Marwan, N.; Kurths, J.; and Donner, R.~V. (community 24);
Riley, M.~A.; Shockley, K.; Turvey, M.~T. (community 32);
Zbilut, J.~P.; Giuliani, A.; Webber, Jr., C.~L. (community 16); and
Wallot, S.; Cox, R.~F.~A.; Hasselman, F. (community 937).
Most of the larger communities are mutually
connected (probably due to more general methodological joint work), but others are more separated, e.g.,
communities 937 (Wallot and Cox et al.), 380 (Litak et al.), and 950 (Acharya et al.), probably because of
some kind of topical specialisation. Nevertheless, the co-author network shows an impressive 
and close exchange of the different scientific fields, and a collaborative community.
(A version of the co-author network including all authors with at least one co-author,
i.e., at least with a network degree of one, is accessible in the Zenodo 
repository, with the link provided in the Data availability section.)


\begin{table}[htbp]
\caption{Largest communities in the co-author network (consisting of at least 1\% of all co-authors), 
together with their first three most productive authors, the size (number of co-authors)
of these communities, and the corresponding fraction on all authors (in \%).}
\begin{center}
\begin{tabular}{c>{\raggedright\arraybackslash}p{6.6cm}cc}
\textbf{Comm.} 	&\textbf{Leading authors} 	&\textbf{Size} 	&\textbf{Fraction}\\
\hline
574	&Wang, Y.; Wang, J.; Wang, X.	&1246	&13.4\\
24	&Marwan, N.; Kurths, J.; Donner, R. V.	&467	&5.0\\
32	&Riley, M. A.; Shockley, K.; Turvey, M. T.	&281	&3.0\\
16	&Zbilut, J. P.; Giuliani, A.; Webber, Jr., C. L.	&255	&2.8\\
937	&Wallot, S.; Cox, R. F. A.; Hasselman, F.	&222	&2.4\\
567	&Yang, H.; Chen, Y.; Bukkapatnam, S. T. S.	&183	&2.0\\
790	&Lee, S.; Park, J.; Singh, A.	&180	&1.9\\
950	&Acharya, U. R.; Faust, O.; Alcaraz, R.	&174	&1.9\\
628	&Li, X.; Soriano, D. C.; Varni, G.	&152	&1.6\\
879	&Lopes, S. R.; Prado, T. D. L.; Viana, R. L.	&144	&1.6\\
380	&Litak, G.; Mosdorf, R.; Syta, A.	&126	&1.4\\
582	&Pham, T. D.; Mukherjee, S.; Li, Q.	&124	&1.3\\
702	&Hirata, Y.; Aihara, K.; Gotoda, H.	&122	&1.3\\
362	&Sujith, R. I.; Sarkar, S.; Pawar, S. A.	&112	&1.2\\
\hline
\end{tabular}
\end{center}
\label{tab_community}
\end{table}

\section{Conclusions}

In this bibliographical study, we have examined the field of recurrence plot
based analysis from different perspectives. While the analysis is grounded in 
simple statistical considerations, more sophisticated analyses could undoubtedly be 
conducted. Nevertheless, it already provides valuable insights into the dynamics 
of the field. There remains ample opportunity for further investigation and 
refinement in the future.

The available data covers almost
40 years. During this time span, we found a rapid growing interest in the
field by a growing number of publications and a growing scientific community.
The growth rate slightly decreased around 2009 (based on number of publications
and size of community), albeit expressing still an exponentially (but slower)
growing field. After 2019, we found another breaking point, which may indicate 
saturation or the reaching of a limit in the expansion of the method to different 
fields and scientific communities.

The major important journals in the field are mainly from physics, covering
interdisciplinary physics, nonlinear dynamics, and chaos. But we also find
some more general journals and high-ranked journals publishing studies based
on recurrence analysis.

A cluster analysis has identified the scientific disciplines where recurrence
plot based methods have found applications. These range from nonlinear time
series analysis and dynamical systems theory, over several life science 
disciplines (cardiovascular, brain, cognitive, social sciences) and engineering
(monitoring, damage detection, combustion optimisation),
to financial markets and protein structure analysis. A remarkable new and quickly 
growing field is the combination of machine learning and recurrence plots for 
classification purposes. The analysis of these fields over time reveals their 
emergence and, for some fields, subsequent (minor) decline.

The citation analysis sheds light on the general citation dynamics. 
Papers are mostly cited after two to four years, but then gradually losing
interest. The peaks in the number maximum citations over time indicate the 
milestones in the development of the recurrence based methods. Citation analysis
also helps to identify the most active groups in the field.

Finally, collaborations on country-level and co-author-level are considered.
This analysis reveals the main institutional and disciplinary communities
active in the field. Moreover, it demonstrates an impressive 
and close exchange of the different scientific fields, and a highly 
collaborative community.

The field on recurrence plots, recurrence networks, and recurrence quantification
analysis is lively and steadily growing. 
The rapidly expanding subject of machine learning is poised to become a major direction 
in the coming years.

\section{Data availability}

The used BibTeX file with the recurrence plot bibliography (as of February 2025
update) and scripts for data retrieval and analysis are available at Zenodo:
\url{https://doi.org/10.5281/zenodo.14904816}.

\clearpage
%
%
%
\bibliographystyle{unsrtnat}

\bibliography{rp,misc}

\begin{thebibliography}{63}
\providecommand{\natexlab}[1]{#1}
\providecommand{\url}[1]{\texttt{#1}}
\expandafter\ifx\csname urlstyle\endcsname\relax
  \providecommand{\doi}[1]{doi: #1}\else
  \providecommand{\doi}{doi: \begingroup \urlstyle{rm}\Url}\fi

\bibitem[Eckmann et~al.(1987)Eckmann, {Oliffson Kamphorst}, and
  Ruelle]{eckmann87}
J.-P. Eckmann, S.~{Oliffson Kamphorst}, and D.~Ruelle.
\newblock {Recurrence Plots} of {Dynamical Systems}.
\newblock \emph{Europhysics Letters}, 4\penalty0 (9):\penalty0 973--977, 1987.
\newblock \doi{10.1209/0295-5075/4/9/004}.
\newblock \ding{115}.

\bibitem[Marwan et~al.(2007)Marwan, Romano, Thiel, and Kurths]{marwan2007}
N.~Marwan, M.~C. Romano, M.~Thiel, and J.~Kurths.
\newblock {Recurrence Plots for the Analysis of Complex Systems}.
\newblock \emph{Physics Reports}, 438\penalty0 (5--6):\penalty0 237--329, 2007.
\newblock \doi{10.1016/j.physrep.2006.11.001}.
\newblock \ding{115}.

\bibitem[Marwan(2008)]{marwan2008epjst}
N.~Marwan.
\newblock {A Historical Review of Recurrence Plots}.
\newblock \emph{European Physical Journal -- Special Topics}, 164\penalty0
  (1):\penalty0 3--12, 2008.
\newblock \doi{10.1140/epjst/e2008-00829-1}.

\bibitem[Zbilut and {Webber, Jr.}(1992)]{zbilut92}
J.~P. Zbilut and C.~L. {Webber, Jr.}
\newblock Embeddings and delays as derived from quantification of recurrence
  plots.
\newblock \emph{Physics Letters A}, 171\penalty0 (3--4):\penalty0 199--203,
  1992.
\newblock \doi{10.1016/0375-9601(92)90426-M}.
\newblock \ding{115}.

\bibitem[{Webber, Jr.} and Zbilut(1994)]{webber94}
C.~L. {Webber, Jr.} and J.~P. Zbilut.
\newblock Dynamical assessment of physiological systems and states using
  recurrence plot strategies.
\newblock \emph{Journal of Applied Physiology}, 76\penalty0 (2):\penalty0
  965--973, 1994.
\newblock \doi{10.1152/jappl.1994.76.2.965}.
\newblock \ding{115}.

\bibitem[Marwan et~al.(2002)Marwan, Wessel, Meyerfeldt, Schirdewan, and
  Kurths]{marwan2002herz}
N.~Marwan, N.~Wessel, U.~Meyerfeldt, A.~Schirdewan, and J.~Kurths.
\newblock {Recurrence Plot Based Measures of Complexity and its Application to
  Heart Rate Variability Data}.
\newblock \emph{{Physical Review E}}, 66\penalty0 (2):\penalty0 026702, 2002.
\newblock \doi{10.1103/PhysRevE.66.026702}.
\newblock \ding{115}.

\bibitem[Marwan et~al.(2009)Marwan, Donges, Zou, Donner, and
  Kurths]{marwan2009b}
N.~Marwan, J.~F. Donges, Y.~Zou, R.~V. Donner, and J.~Kurths.
\newblock Complex network approach for recurrence analysis of time series.
\newblock \emph{Physics Letters A}, 373\penalty0 (46):\penalty0 4246--4254,
  2009.
\newblock \doi{10.1016/j.physleta.2009.09.042}.

\bibitem[Donner et~al.(2010{\natexlab{a}})Donner, Zou, Donges, Marwan, and
  Kurths]{donner2010b}
R.~V. Donner, Y.~Zou, J.~F. Donges, N.~Marwan, and J.~Kurths.
\newblock {Recurrence networks -- A novel paradigm for nonlinear time series
  analysis}.
\newblock \emph{New Journal of Physics}, 12\penalty0 (3):\penalty0 033025,
  2010{\natexlab{a}}.
\newblock \doi{10.1088/1367-2630/12/3/033025}.
\newblock Listed as <a
  href="https://iopscience.iop.org/journal/1367-2630/page/Best
  of 2010 in the New Journal of Physics</a>.

\bibitem[Zou et~al.(2019)Zou, Donner, Marwan, Donges, and Kurths]{zou2019}
Y.~Zou, R.~V. Donner, N.~Marwan, J.~F. Donges, and J.~Kurths.
\newblock Complex network approaches to nonlinear time series analysis.
\newblock \emph{Physics Reports}, 787:\penalty0 1--97, 2019.
\newblock \doi{10.1016/j.physrep.2018.10.005}.

\bibitem[Little et~al.(2007)Little, {McSharry}, Roberts, Costello, and
  Moroz]{little2007}
M.~A. Little, P.~E. {McSharry}, S.~J. Roberts, D.~A.~E. Costello, and I.~M.
  Moroz.
\newblock {Exploiting Nonlinear Recurrence and Fractal Scaling Properties for
  Voice Disorder Detection}.
\newblock \emph{Biomedical Engineering Online}, 6\penalty0 (23):\penalty0
  1--19, 2007.
\newblock \doi{10.1186/1475-925X-6-23}.

\bibitem[Marwan and Kraemer(2023)]{marwan2023}
N.~Marwan and K.~H. Kraemer.
\newblock Trends in recurrence analysis of dynamical systems.
\newblock \emph{European Physical Journal -- Special Topics}, 232:\penalty0
  5--27, 2023.
\newblock \doi{10.1140/epjs/s11734-022-00739-8}.

\bibitem[Wessel et~al.(2001)Wessel, Marwan, Meyerfeldt, Schirdewan, and
  Kurths]{wessel2001}
N.~Wessel, N.~Marwan, U.~Meyerfeldt, A.~Schirdewan, and J.~Kurths.
\newblock Recurrence quantification analysis to characterise the heart rate
  variability before the onset of ventricular tachycardia.
\newblock \emph{Lecture Notes in Computer Science}, 2199:\penalty0 295--301,
  2001.
\newblock \doi{10.1007/3-540-45497-7_45}.

\bibitem[Iwaniec and Iwaniec(2018)]{iwaniec2018}
J.~Iwaniec and M.~Iwaniec.
\newblock {Application of Recurrence-Based Methods to Heart Work Analysis}.
\newblock \emph{Applied Condition Monitoring}, 10:\penalty0 343--352, 2018.
\newblock \doi{10.1007/978-3-319-62042-8_31}.

\bibitem[{Calder\'on-Ju\'arez} et~al.(2023){Calder\'on-Ju\'arez},
  {Gonz\'alez-G\'omez}, Echeverr{\'\i}a, and Lerma]{calderonjuarez2023d}
M.~{Calder\'on-Ju\'arez}, G.~H. {Gonz\'alez-G\'omez}, J.~Echeverr{\'\i}a, and
  C.~Lerma.
\newblock {Revisiting nonlinearity of heart rate variability in healthy aging}.
\newblock \emph{Scientific Reports}, 13\penalty0 (1):\penalty0 13185, 2023.
\newblock \doi{10.1038/s41598-023-40385-1}.

\bibitem[Acharya et~al.(2011)Acharya, Chua, Faust, Lim, and Lim]{acharya2011a}
U.~R. Acharya, E.~C.~P. Chua, O.~Faust, T.-C. Lim, and L.~F.~B. Lim.
\newblock Automated detection of sleep apnea from electrocardiogram signals
  using nonlinear parameters.
\newblock \emph{Physiological Measurement}, 32\penalty0 (3):\penalty0 287--303,
  2011.
\newblock \doi{10.1088/0967-3334/32/3/002}.

\bibitem[Ngamga et~al.(2016)Ngamga, Bialonski, Marwan, Kurths, Geier, and
  Lehnertz]{ngamga2016}
E.~J. Ngamga, S.~Bialonski, N.~Marwan, J.~Kurths, C.~Geier, and K.~Lehnertz.
\newblock {Evaluation of selected recurrence measures in discriminating
  pre-ictal and inter-ictal periods from epileptic EEG data}.
\newblock \emph{Physics Letters A}, 380\penalty0 (16):\penalty0 1419--1425,
  2016.
\newblock \doi{10.1016/j.physleta.2016.02.024}.

\bibitem[Billeci et~al.(2018)Billeci, Marino, Insana, Vatti, and
  Varanini]{billeci2018b}
L.~Billeci, D.~Marino, L.~Insana, G.~Vatti, and M.~Varanini.
\newblock {Patient-specific seizure prediction based on heart rate variability
  and recurrence quantification analysis}.
\newblock \emph{PLoS ONE}, 13\penalty0 (9):\penalty0 0204339, 2018.
\newblock \doi{10.1371/journal.pone.0204339}.

\bibitem[Riley and Clark(2003)]{riley2003}
M.~A. Riley and S.~Clark.
\newblock Recurrence analysis of human postural sway during the sensory
  organization test.
\newblock \emph{Neuroscience Letters}, 342\penalty0 (1--2):\penalty0 45--48,
  2003.
\newblock \doi{10.1016/S0304-3940(03)00229-5}.

\bibitem[Wallot et~al.(2013)Wallot, Hollis, and {van Rooij}]{wallot2013}
S.~Wallot, G.~Hollis, and M.~{van Rooij}.
\newblock {Connected Text Reading and Differences in Text Reading Fluency in
  Adult Readers}.
\newblock \emph{PLoS ONE}, 8\penalty0 (8):\penalty0 e71914, 2013.
\newblock \doi{10.1371/journal.pone.0071914}.

\bibitem[{beim Graben} and Hutt(2014)]{beimgraben2014}
P.~{beim Graben} and A.~Hutt.
\newblock {Detecting event-related recurrences by symbolic analysis:
  Applications to human language processing}.
\newblock \emph{Philosophical Transactions of the Royal Society A},
  373\penalty0 (2034):\penalty0 20140089, 2014.
\newblock \doi{10.1098/rsta.2014.0089}.

\bibitem[Vlahogianni et~al.(2008)Vlahogianni, Karlaftis, and
  Golias]{vlahogianni2008a}
E.~I. Vlahogianni, M.~G. Karlaftis, and J.~C. Golias.
\newblock {Temporal Evolution of Short-Term Urban Traffic Flow: A Nonlinear
  Dynamics Approach}.
\newblock \emph{Computer-Aided Civil and Infrastructure Engineering},
  23\penalty0 (7):\penalty0 536--548, 2008.
\newblock \doi{10.1111/j.1467-8667.2008.00554.x}.

\bibitem[Mosdorf(2012)]{mosdorf2012}
M.~Mosdorf.
\newblock Method for detecting software anomalies based on recurrence plot
  analysis.
\newblock \emph{Journal of Theoretical and Applied Computer Science},
  6\penalty0 (1):\penalty0 3--12, 2012.
\newblock URL \url{http://www.jtacs.org/archive/2012/1/1}.

\bibitem[Godavarthi et~al.(2018)Godavarthi, Pawar, Unni, Sujith, Marwan, and
  Kurths]{godavarthi2018}
V.~Godavarthi, S.~A. Pawar, V.~R. Unni, R.~I. Sujith, N.~Marwan, and J.~Kurths.
\newblock Coupled interaction between unsteady flame dynamics and acoustic
  field in a turbulent combustor.
\newblock \emph{Chaos}, 28:\penalty0 113111, 2018.
\newblock \doi{10.1063/1.5052210}.

\bibitem[Stender et~al.(2019)Stender, Oberst, Tiedemann, and
  Hoffmann]{stender2019b}
M.~Stender, S.~Oberst, M.~Tiedemann, and N.~P. Hoffmann.
\newblock {Complex machine dynamics: systematic recurrence quantification
  analysis of disk brake vibration data}.
\newblock \emph{Nonlinear Dynamics}, 97:\penalty0 2483--2497, 2019.
\newblock \doi{10.1007/s11071-019-05143-x}.

\bibitem[Syta et~al.(2023)Syta, Czarnigowski, Jakli\'{n}ski, and
  Marwan]{syta2023}
A.~Syta, J.~Czarnigowski, P.~Jakli\'{n}ski, and N.~Marwan.
\newblock {Detection and identification of cylinder misfire in small aircraft
  engine in different operating conditions by linear and non-linear properties
  of frequency components}.
\newblock \emph{Measurement}, 223:\penalty0 113763, 2023.
\newblock \doi{10.1016/j.measurement.2023.113763}.

\bibitem[{\v C}erm\'ak et~al.(2008){\v C}erm\'ak, {\v S}afanda, and
  Bodri]{cermak2008b}
V.~{\v C}erm\'ak, J.~{\v S}afanda, and L.~Bodri.
\newblock {Precise temperature monitoring in boreholes: evidence for
  oscillatory convection? Part 1: Experiments and field data}.
\newblock \emph{International Journal of Earth Sciences}, 97\penalty0
  (2):\penalty0 365--373, 2008.
\newblock \doi{10.1007/s00531-007-0237-4}.

\bibitem[Eroglu et~al.(2016)Eroglu, {McRobie}, Ozken, Stemler, Wyrwoll,
  Breitenbach, Marwan, and Kurths]{eroglu2016}
D.~Eroglu, F.~H. {McRobie}, I.~Ozken, T.~Stemler, K.-H. Wyrwoll, S.~F.~M.
  Breitenbach, N.~Marwan, and J.~Kurths.
\newblock {See-saw relationship of the Holocene East Asian-Australian summer
  monsoon}.
\newblock \emph{Nature Communications}, 7:\penalty0 12929, 2016.
\newblock \doi{10.1038/ncomms12929}.

\bibitem[Oberst et~al.(2018)Oberst, Niven, Lester, Ord, Hobbs, and
  Hoffmann]{oberst2018b}
S.~Oberst, R.~K. Niven, D.~R. Lester, A.~Ord, B.~Hobbs, and N.~P. Hoffmann.
\newblock {Detection of unstable periodic orbits in mineralising geological
  systems}.
\newblock \emph{Chaos}, 28\penalty0 (8):\penalty0 085711, 2018.
\newblock \doi{10.1063/1.5024134}.

\bibitem[Braun et~al.(2023)Braun, Kraemer, and Marwan]{braun2023}
T.~Braun, K.~H. Kraemer, and N.~Marwan.
\newblock {Recurrence flow measure of nonlinear dependence}.
\newblock \emph{European Physical Journal -- Special Topics}, 232:\penalty0
  57--67, 2023.
\newblock \doi{10.1140/epjs/s11734-022-00687-3}.

\bibitem[rpw(2022)]{rpwebsite_bibliography}
{Recurrence Plots and Cross Recurrence Plots: A Comprehensive Bibliography
  About RPs, RQA And Their Applications}.
\newblock \url{http://www.recurrence-plot.tk/bibliography.php}, 2022.

\bibitem[Singh et~al.(2022)Singh, Barme, Ward, Tupikina, and
  Santolini]{singh2022bibliographic}
C.~K. Singh, E.~Barme, R.~Ward, L.~Tupikina, and M.~Santolini.
\newblock {Quantifying the rise and fall of scientific fields}.
\newblock \emph{PLOS ONE}, 17\penalty0 (6):\penalty0 e0270131, 2022.
\newblock \doi{10.1371/journal.pone.0270131}.

\bibitem[way(2003)]{wayback2003}
{Internet Archive (WaybackMachine): A COMPREHENSIVE BIBLIOGRAPHY ABOUT RPs, RQA
  AND THEIR APPLICATIONS}.
\newblock
  \url{https://web.archive.org/web/20030717152703/https://agnld.uni-potsdam.de/~marwan/literature.php},
  2003.

\bibitem[Trulla et~al.(1996)Trulla, Giuliani, Zbilut, and {Webber,
  Jr.}]{trulla96}
L.~L. Trulla, A.~Giuliani, J.~P. Zbilut, and C.~L. {Webber, Jr.}
\newblock Recurrence quantification analysis of the logistic equation with
  transients.
\newblock \emph{Physics Letters A}, 223\penalty0 (4):\penalty0 255--260, 1996.
\newblock \doi{10.1016/S0375-9601(96)00741-4}.

\bibitem[Zbilut et~al.(1998)Zbilut, Giuliani, and {Webber, Jr.}]{zbilut98}
J.~P. Zbilut, A.~Giuliani, and C.~L. {Webber, Jr.}
\newblock Detecting deterministic signals in exceptionally noisy environments
  using cross-recurrence quantification.
\newblock \emph{Physics Letters A}, 246\penalty0 (1--2):\penalty0 122--128,
  1998.
\newblock \doi{10.1016/S0375-9601(98)00457-5}.

\bibitem[Romano et~al.(2004)Romano, Thiel, Kurths, and {von Bloh}]{romano2004}
M.~C. Romano, M.~Thiel, J.~Kurths, and W.~{von Bloh}.
\newblock {Multivariate Recurrence Plots}.
\newblock \emph{Physics Letters A}, 330\penalty0 (3--4):\penalty0 214--223,
  2004.
\newblock \doi{10.1016/j.physleta.2004.07.066}.
\newblock \ding{115}.

\bibitem[Marwan and Kurths(2005)]{marwan2005}
N.~Marwan and J.~Kurths.
\newblock Line structures in recurrence plots.
\newblock \emph{Physics Letters A}, 336\penalty0 (4--5):\penalty0 349--357,
  2005.
\newblock \doi{10.1016/j.physleta.2004.12.056}.
\newblock \ding{115}.

\bibitem[Facchini et~al.(2005)Facchini, Kantz, and Tiezzi]{facchini2005}
A.~Facchini, H.~Kantz, and E.~B.~P. Tiezzi.
\newblock {Recurrence plot analysis of nonstationary data: The understanding of
  curved patterns}.
\newblock \emph{Physical Review E}, 72:\penalty0 021915, 2005.
\newblock \doi{10.1103/PhysRevE.72.021915}.

\bibitem[Angus et~al.(2012)Angus, Smith, and Wiles]{angus2012}
D.~Angus, A.~E. Smith, and J.~Wiles.
\newblock {Conceptual Recurrence Plots: Revealing Patterns in Human Discourse}.
\newblock \emph{IEEE Transactions on Visualization and Computer Graphics},
  18\penalty0 (6):\penalty0 988--997, 2012.
\newblock \doi{10.1109/TVCG.2011.100}.

\bibitem[Yang and Chen(2014)]{yang2014}
H.~Yang and Y.~Chen.
\newblock Heterogeneous recurrence monitoring and control of nonlinear
  stochastic processes.
\newblock \emph{Chaos}, 24:\penalty0 013138, 2014.
\newblock \doi{10.1063/1.4869306}.

\bibitem[Corso et~al.(2018)Corso, Prado, {dos Santos Lima}, Kurths, and
  Lopes]{corso2018}
G.~Corso, T.~D.~L. Prado, G.~Z. {dos Santos Lima}, J.~Kurths, and S.~R. Lopes.
\newblock {Quantifying entropy using recurrence matrix microstates}.
\newblock \emph{Chaos}, 28\penalty0 (8):\penalty0 083108, 2018.
\newblock \doi{10.1063/1.5042026}.

\bibitem[Braun et~al.(2021)Braun, Unni, Sujith, Kurths, and Marwan]{braun2021}
T.~Braun, V.~R. Unni, R.~I. Sujith, J.~Kurths, and N.~Marwan.
\newblock {Detection of dynamical regime transitions with lacunarity as a
  multiscale recurrence quantification measure}.
\newblock \emph{Nonlinear Dynamics}, 104:\penalty0 3955--3973, 2021.
\newblock \doi{10.1007/s11071-021-06457-5}.

\bibitem[Hirata et~al.(2021)Hirata, Kitanishi, Sugishita, and
  Gotoh]{hirata2021a}
Y.~Hirata, Y.~Kitanishi, H.~Sugishita, and Y.~Gotoh.
\newblock {Fast reconstruction of an original continuous series from a
  recurrence plot}.
\newblock \emph{Chaos}, 31\penalty0 (12):\penalty0 121101, 2021.
\newblock \doi{10.1063/5.0073899}.

\bibitem[Ramdani et~al.(2016)Ramdani, Bouchara, Lagarde, and
  Lesne]{ramdani2016}
S.~Ramdani, F.~Bouchara, J.~Lagarde, and A.~Lesne.
\newblock {Recurrence plots of discrete-time Gaussian stochastic processes}.
\newblock \emph{Physica A}, 330:\penalty0 17--31, 2016.
\newblock \doi{10.1016/j.physd.2016.04.017}.

\bibitem[{\v S}pitalsk{\'y}(2018)]{spitalsky2018}
V.~{\v S}pitalsk{\'y}.
\newblock {Local correlation entropy}.
\newblock \emph{Discrete and Continuous Dynamical Systems}, 38\penalty0
  (11):\penalty0 5711--5733, 2018.
\newblock \doi{10.3934/dcds.2018249}.

\bibitem[Medrano et~al.(2021)Medrano, Kheddar, Lesne, and Ramdani]{medrano2021}
J.~Medrano, A.~Kheddar, A.~Lesne, and S.~Ramdani.
\newblock {Radius selection using kernel density estimation for the computation
  of nonlinear measures}.
\newblock \emph{Chaos}, 31\penalty0 (8):\penalty0 083131, 2021.
\newblock \doi{10.1063/5.0055797}.

\bibitem[Hirata and Shiro(2023)]{hirata2023a}
Y.~Hirata and M.~Shiro.
\newblock {Recurrence plots bridge deterministic systems and stochastic systems
  topologically and measure-theoretically}.
\newblock \emph{Chaos}, 33\penalty0 (8):\penalty0 083118, 2023.
\newblock \doi{10.1063/5.0156945}.

\bibitem[Davies and Bouldin(1979)]{davies1979}
D.~L. Davies and D.~W. Bouldin.
\newblock {A Cluster Separation Measure}.
\newblock \emph{IEEE Transactions on Pattern Analysis and Machine
  Intelligence}, PAMI-1\penalty0 (2):\penalty0 224--227, 1979.
\newblock \doi{10.1109/TPAMI.1979.4766909}.

\bibitem[Rousseeuw(1987)]{rousseeuw1987}
Peter~J. Rousseeuw.
\newblock {Silhouettes: A graphical aid to the interpretation and validation of
  cluster analysis}.
\newblock \emph{Journal of Computational and Applied Mathematics}, 20:\penalty0
  53--65, 1987.
\newblock \doi{10.1016/0377-0427(87)90125-7}.

\bibitem[Portenoy et~al.(2017)Portenoy, Hullman, and West]{portenoy2017}
J.~Portenoy, J.~Hullman, and J.~D. West.
\newblock {Leveraging Citation Networks to Visualize Scholarly Influence Over
  Time}.
\newblock \emph{Frontiers in Research Metrics and Analytics}, 2:\penalty0 8,
  2017.
\newblock \doi{10.3389/frma.2017.00008}.

\bibitem[Yan and Ding(2010)]{yan2010a}
E.~Yan and Y.~Ding.
\newblock {Weighted citation: An indicator of an article's prestige}.
\newblock \emph{Journal of the American Society for Information Science and
  Technology}, 61\penalty0 (8):\penalty0 1635--1643, 2010.
\newblock \doi{10.1002/asi.21349}.

\bibitem[{Webber, Jr.} and Marwan(2015)]{webber2015}
C.~L. {Webber, Jr.} and N.~Marwan.
\newblock \emph{{Recurrence Quantification Analysis -- Theory and Best
  Practices}}.
\newblock Springer, Cham, 2015.
\newblock ISBN 978-3-319-07154-1.
\newblock \doi{10.1007/978-3-319-07155-8}.

\bibitem[Donner et~al.(2010{\natexlab{b}})Donner, Zou, Donges, Marwan, and
  Kurths]{donner2010a}
R.~V. Donner, Y.~Zou, J.~F. Donges, N.~Marwan, and J.~Kurths.
\newblock Ambiguities in recurrence-based complex network representations of
  time series.
\newblock \emph{Physical Review E}, 81:\penalty0 015101(R), 2010{\natexlab{b}}.
\newblock \doi{10.1103/PhysRevE.81.015101}.

\bibitem[Donner et~al.(2011)Donner, Small, Donges, Marwan, Zou, Xiang, and
  Kurths]{donner2011}
R.~V. Donner, M.~Small, J.~F. Donges, N.~Marwan, Y.~Zou, R.~Xiang, and
  J.~Kurths.
\newblock Recurrence-based time series analysis by means of complex network
  methods.
\newblock \emph{International Journal of Bifurcation and Chaos}, 21\penalty0
  (4):\penalty0 1019--1046, 2011.
\newblock \doi{10.1142/S0218127411029021}.

\bibitem[Donges et~al.(2012)Donges, Heitzig, Donner, and Kurths]{donges2012}
J.~F. Donges, J.~Heitzig, R.~V. Donner, and J.~Kurths.
\newblock Analytical framework for recurrence network analysis of time series.
\newblock \emph{Physical Review E}, 85:\penalty0 046105, 2012.
\newblock \doi{10.1103/PhysRevE.85.046105}.

\bibitem[Mathunjwa et~al.(2021)Mathunjwa, Lin, Lin, Abbod, and
  Shieh]{mathunjwa2021}
B.~Mathunjwa, Y.~Lin, C.~Lin, M.~Abbod, and J.~Shieh.
\newblock {ECG arrhythmia classification by using a recurrence plot and
  convolutional neural network}.
\newblock \emph{Biomedical Signal Processing and Control}, 64:\penalty0 102262,
  2021.
\newblock \doi{10.1016/j.bspc.2020.102262}.

\bibitem[Zhang et~al.(2022)Zhang, Hou, OuYang, and Zhou]{zhang2022}
Y.~Zhang, Y.~Hou, K.~OuYang, and S.~Zhou.
\newblock {Multi-scale signed recurrence plot based time series classification
  using inception architectural networks}.
\newblock \emph{Pattern Recognition}, 123:\penalty0 108385, 2022.
\newblock \doi{10.1016/j.patcog.2021.108385}.

\bibitem[Marwan and Kurths(2002)]{marwan2002pla}
N.~Marwan and J.~Kurths.
\newblock {Nonlinear analysis of bivariate data with cross recurrence plots}.
\newblock \emph{{Physics Letters A}}, 302\penalty0 (5--6):\penalty0 299--307,
  2002.
\newblock \doi{10.1016/S0375-9601(02)01170-2}.
\newblock \ding{115}.

\bibitem[Marwan(2011)]{marwan2011}
N.~Marwan.
\newblock How to avoid potential pitfalls in recurrence plot based data
  analysis.
\newblock \emph{International Journal of Bifurcation and Chaos}, 21\penalty0
  (4):\penalty0 1003--1017, 2011.
\newblock \doi{10.1142/S0218127411029008}.

\bibitem[Thiel et~al.(2004)Thiel, Romano, Read, and Kurths]{thiel2004a}
M.~Thiel, M.~C. Romano, P.~L. Read, and J.~Kurths.
\newblock {Estimation of dynamical invariants without embedding by recurrence
  plots}.
\newblock \emph{Chaos}, 14\penalty0 (2):\penalty0 234--243, 2004.
\newblock \doi{10.1063/1.1667633}.
\newblock \ding{115}.

\bibitem[Romano et~al.(2005)Romano, Thiel, Kurths, Kiss, and
  Hudson]{romano2005}
M.~C. Romano, M.~Thiel, J.~Kurths, I.~Z. Kiss, and J.~L. Hudson.
\newblock Detection of synchronization for non-phase-coherent and
  non-stationary data.
\newblock \emph{Europhysics Letters}, 71\penalty0 (3):\penalty0 466--472, 2005.
\newblock \doi{10.1209/epl/i2005-10095-1}.
\newblock \ding{115}.

\bibitem[Hatami et~al.(2018)Hatami, Gavet, and Debayle]{hatami2018}
N.~Hatami, Y.~Gavet, and J.~Debayle.
\newblock {Classification of time-series images using deep convolutional neural
  networks}.
\newblock \emph{Proceedings of SPIE}, 10696:\penalty0 106960Y, 2018.
\newblock \doi{10.1117/12.2309486}.

\bibitem[Asghari et~al.(2004)Asghari, Broeg, Carone, Casas-Miranda, Palacio,
  Csillik, Dvorak, Freistetter, Hadjivantsides, Hussmann, Khramova,
  Khristoforova, Khromova, Kitiashivilli, Kozlowski, Laakso, Laczkowski,
  Lytvinenko, Miloni, Morishima, Moro-Martin, Paksyutov, Pal, Patidar, Pecnik,
  Peles, Pyo, Quinn, Rodriguez, Romano, Saikia, Stadel, Thiel, Todorovic,
  Veras, {Vieira Neto}, Vilagi, {von Bloh}, Zechner, and
  Zhuchkova]{asghari2004}
N.~Asghari, C.~Broeg, L.~Carone, R.~Casas-Miranda, J.~C.~Castro Palacio,
  I.~Csillik, R.~Dvorak, F.~Freistetter, G.~Hadjivantsides, H.~Hussmann,
  A.~Khramova, M.~Khristoforova, I.~Khromova, I.~Kitiashivilli, S.~Kozlowski,
  T.~Laakso, T.~Laczkowski, D.~Lytvinenko, O.~Miloni, R.~Morishima,
  A.~Moro-Martin, V.~Paksyutov, A.~Pal, V.~Patidar, B.~Pecnik, O.~Peles,
  J.~Pyo, T.~Quinn, A.~Rodriguez, M.~C. Romano, E.~Saikia, J.~Stadel, M.~Thiel,
  N.~Todorovic, D.~Veras, E.~{Vieira Neto}, J.~Vilagi, W.~{von Bloh},
  R.~Zechner, and E.~Zhuchkova.
\newblock Stability of terrestrial planets in the habitable zone of gl 777 a,
  hd 72659, gl 614, 47 uma and hd 4208.
\newblock \emph{Astronomy \& Astrophysics}, 426:\penalty0 353--365, 2004.
\newblock \doi{10.1051/0004-6361:20040390}.

\bibitem[Blondel et~al.(2008)Blondel, Guillaume, Lambiotte, and
  Lefebvre]{blondel2008}
V.~D. Blondel, J.-L. Guillaume, R.~Lambiotte, and E.~Lefebvre.
\newblock {Fast unfolding of communities in large networks}.
\newblock \emph{Journal of Statistical Mechanics: Theory and Experiment},
  2008\penalty0 (10):\penalty0 P10008, 2008.
\newblock \doi{10.1088/1742-5468/2008/10/P10008}.

\end{thebibliography}

\end{document}